\documentclass[fleqn,10pt]{wlscirep}
\usepackage[utf8]{inputenc}
\usepackage[T1]{fontenc}

\usepackage{footnote}

\usepackage{array}
\renewcommand*{\arraystretch}{1.1}
\newcommand*{\mline}[1]{%
\begingroup
    \renewcommand*{\arraystretch}{1.1}%
   \begin{tabular}[c]{@{}>{\raggedright\arraybackslash}p{2cm}@{}}#1\end{tabular}%
  \endgroup
}

\title{A fixed phase tunable directional coupler based on coupling tuning}

\author[1]{Yang Yang}
\author[1]{Tim Weiss}
\author[1]{Hamed Arianfard}
\author[1]{Akram Youssry}
\author[1,2,*]{Alberto Peruzzo}
\affil[1]{Quantum Photonics Laboratory and Centre for Quantum Computation and Communication Technology, RMIT University, Melbourne, VIC 3000, Australia}
\affil[2]{Qubit Pharmaceuticals, Advanced Research Department, Paris, France}

\affil[*]{alberto.peruzzo@rmit.edu.au}

\begin{abstract}
The field of photonic integrated circuits has witnessed significant progress in recent years, with a growing demand for devices that offer high-performance reconfigurability. 
Due to the inability of conventional tunable directional couplers (TDCs) to maintain a fixed phase while tuning the reflectivity, Mach-Zehnder interferometers (MZIs) are employed as the primary building blocks for reflectivity tuning in constructing large-scale circuits. However, MZIs are prone to fabrication errors due to the need for perfect balanced directional couplers to achieve 0-1 reflectivity, which hinders their scalability.
In this study, we introduce a design of a TDC based on coupling constant tuning in the thin film Lithium Niobate platform and present an optimized design. Our optimized TDC design enables arbitrary reflectivity tuning while ensuring a consistent phase across a wide range of operating wavelengths. Furthermore, it exhibits fewer bending sections than MZIs and is inherently resilient to fabrication errors in waveguide geometry and coupling length compared to both MZIs and conventional TDCs. Our work contributes to developing high-performance photonic integrated circuits with implications for various fields, including optical communication systems and quantum information processing.
\end{abstract}
\begin{document}

\flushbottom
\maketitle
%
%
\thispagestyle{empty}


\section*{Introduction}

Photonic integrated circuits (PICs) have emerged as a promising technology for enabling complex optical functionalities with a small footprint~\cite{chrostowski_hochberg_2015, InP_based_photonic, Zhu:21}. 
One significant advancement in PICs is the development of programmable PICs~\cite{Bogaerts_2020}, which are highly versatile devices capable of processing optical signals in a reconfigurable manner. 
These circuits have exhibited remarkable capabilities finding extensive applications in diverse fields, including radio-frequency signal processing~\cite{Marpaung_2019}, machine learning~\cite{Shen_2017, Harris:18}, condensed matter physics systems simulations~\cite{Yang_2024}, and quantum information processing~\cite{shadbolt2012generating,doi:10.1126/science.aab3642}.

\begin{figure*}
    \centering
    \captionsetup{labelfont={color=black}}
    \includegraphics[width=0.7\columnwidth]{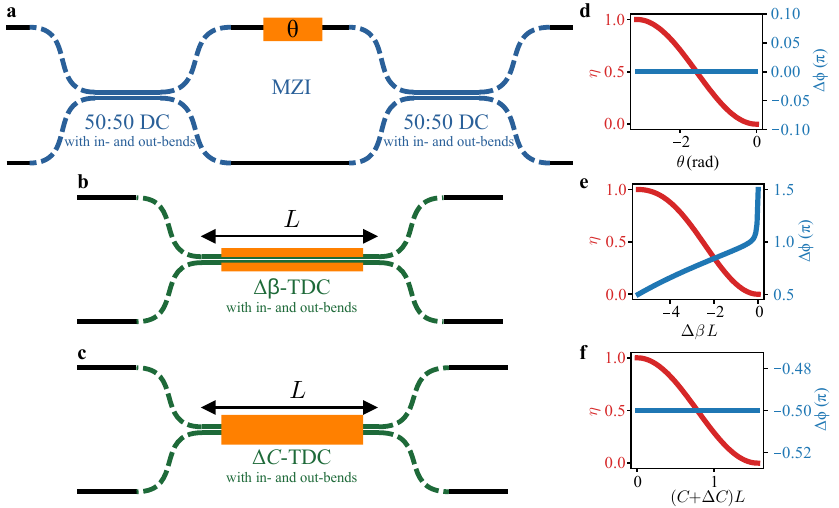}
    \caption{\textbf{Fundamental reflectivity-tuning devices and their phase-reflectivity dependence.} (\textbf{a}) An MZI with a tunable phase shift $\theta$. (\textbf{b}) A conventional $\Delta\beta$-TDC with a coupling length $L$, controlled by propagation constant tuners. (\textbf{c}) The proposed $\Delta{C}$-TDC with a coupling length $L$, controlled by a coupling coefficient tuner. The numerical simulation of the reflectivity $\eta$ (red curve) and relative phase between the optical paths $\Delta \phi$ (blue curve) for the MZI. (\textbf{This work.}) (\textbf{d}), the $\Delta\beta$-TDC (\textbf{e}), and the $\Delta{C}$-TDC (\textbf{f}). The relative phase shift is constant in the tunable range only for the MZI and the proposed $\Delta{C}$-TDC, providing an advantage over the $\Delta\beta$-TDC for phase-sensitive applications.   
    (Numerical simulation details can be found in the supplementary materials.)
    }
    \label{fig:1}
\end{figure*}

To meet the demands of more advanced applications, the integration density of PICs is increasing rapidly~\cite{InP_based_photonic}.
The most adaptable strategies involve constructing waveguide mesh structures or multi-port universal interferometers~\cite{Reck_1994, Clements_2016, Perez_2017} based on the fundamental tunable device that realizes universal 2$\times$2 transformation. This device is implemented by the Mach-Zehnder interferometer (MZI) with an extra phase shifter for relative phase tuning of the output states~\cite{Shen_2017, Harris:18, Harris_2017, doi:10.1126/science.aar7053, Qiang_2018,Bao_2023}. Such an MZI consists of two balanced (50:50) beam splitters and a phase shifter for reflectivity tuning (Fig~\ref{fig:1} (a) and (d)). (Reflectivity $\eta$ is defined as $\eta = 1 - \kappa$, following the convention in the field of quantum optics~\cite{doi:https://doi.org/10.1002/9783527695805.ch5,kok2007linear}, where $\kappa$ represents the coupling or power transfer ratio~\cite{doi:https://doi.org/10.1002/0471213748.ch7,doi:https://doi.org/10.1002/0471213748.ch18}.) However, employing a one-to-one component mapping approach to transferring MZIs from bulk free-space optics to PICs may not be the optimal strategy. This is because the design of MZIs is sensitive to fabrication errors and exhibits more bending losses~\cite{Miller_2015, Burgwal:17}. Achieving balanced directional couplers (DCs) requires perfect fabrication and is crucial for constructing MZIs to achieve 0-1 reflectivity. Additionally, each MZI has 4 bending sections (Fig~\ref{fig:1} (a)), which introduces optical losses and can decrease the fidelity of larger scale circuits~\cite{Clements_2016}. 

\begin{figure*}
    \centering
    \includegraphics[width=0.8\columnwidth]{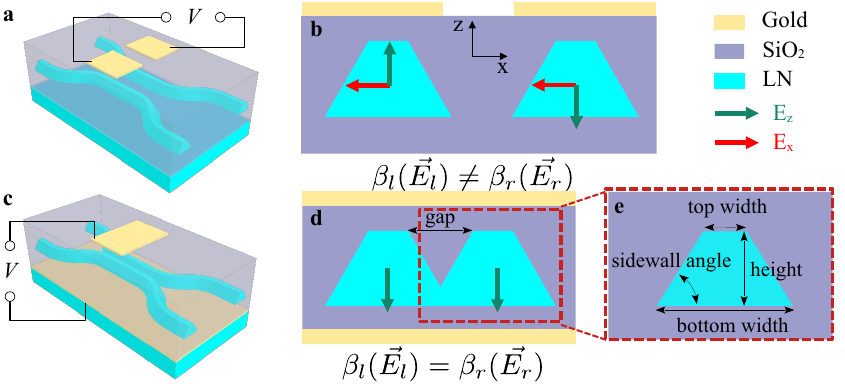}
    \caption{\textbf{Z-cut TFLN TDC schematics.} (\textbf{a}) A schematic of a conventional $\Delta\beta$-TDC example. (\textbf{b}) The cross-section of the $\Delta\beta$-TDC is shown in (\textbf{a}). The inset coordinate indicates the LN crystal orientation. Green and red arrows indicate the electric field direction at the center of each waveguide when a voltage is applied. By applying a voltage to the right electrode, the left and right waveguides experience the electric field in a different direction ($\vec{E_l}$ and $\vec{E_r}$), creating a difference between the propagation constants of the left and right waveguides is created ($\Delta\beta=|\beta_l(\vec{E_l})-\beta_r(\vec{E_r})|$).  Propagation with different propagation constants enables tunability and leads directly to a buildup of a relative phase. (\textbf{c}) Proposed $\Delta{C}$-TDC schematics (optimized geometry parameters are reported in Fig~\ref{fig:4}). (\textbf{d}) The cross-section of the coupled region. The green arrow indicates the electric field direction at the center of each waveguide when a voltage is applied to the top electrode. Because both the left and right waveguides experience the same electric field, no propagation constant difference between the left and right waveguides is introduced ($\Delta\beta=|\beta_l(\vec{E_l})-\beta_r(\vec{E_r})|=0$). (\textbf{e}) Cross-sectional view and descriptive parameters of an individual waveguide within the coupler structure.}
    \label{fig:2}
\end{figure*}

As an alternative to MZIs, the conventional $\Delta\beta$-based TDC ($\Delta\beta$-TDC) (see the illustration in Fig~\ref{fig:1} (b) and an example device schematic in Fig~\ref{fig:2} (a) and (b)), which relies on creating a difference in propagation constants between the two waveguides to tune the reflectivity, have less bending sections. However, such a device is wavelength-dependent and sensitive to fabrication error. Furthermore, both its reflectivity and phase shift are dependent on the voltage, and cannot be independently controlled~\cite{doi:https://doi.org/10.1002/0471213748.ch18, 826874, Okayama} (Fig~\ref{fig:1} (e)). Moreover, the exact phase-voltage dependence has to be characterized after the fabrication of the device, which is experimentally challenging for high-dimensional photonic circuits, or circuits consisting of cascaded sections. In applications where phase shifts are critical (such as in quantum information processing), this will cause a degradation in the overall performance of the system.

In this paper, we propose a $\Delta{C}$-based TDC ($\Delta{C}$-TDC) (see the illustration in Fig~\ref{fig:1} (c) and the device schematic Fig~\ref{fig:2} (c)) that implements the same unitary transformation as an MZI, but possesses fewer bending sections and enhanced fabrication error tolerance, with a method to assess its performance. This addresses the scalability challenges, potentially enabling more advanced applications associated with using MZIs. Unlike conventional $\Delta\beta$-TDCS, our $\Delta{C}$-TDC design achieves tunability through coupling tuning, which maintains the relative phase between coupled waveguides fixed (Fig~\ref{fig:1} (e) and (f)) and is significantly less sensitive to coupling length and operating wavelength. Moreover, our design eliminates the issue of non-even responses from controllers in the dual-drive design~\cite{8351983,Perez-Lopez:19}, ensuring consistent and reliable performance over an extra-wide range of wavelengths. 

\section*{Results and methods}

\begin{figure}
    \centering
    \includegraphics[width=0.8\columnwidth]{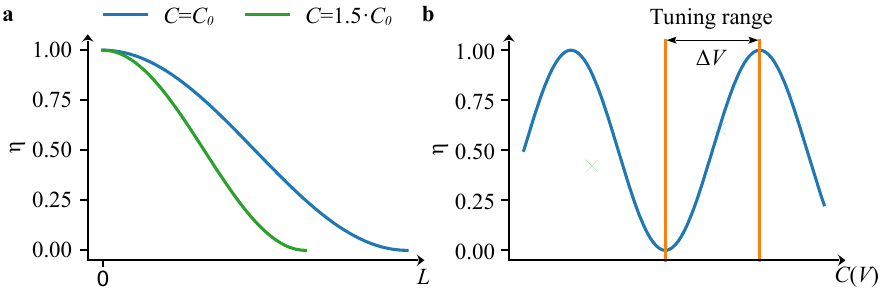}
    \caption{\textbf{$\Delta{C}$-TDC tuning principle.} (\textbf{a}) Reflectivity curves of DCs with different coupling coefficients along the propagation of the coupled region ($C_0$ is an arbitrary value). (\textbf{b}) Output reflectivity change of a $\Delta{C}$-TDC, which is achieved by controlling the coupling coefficient electro-optically. Note: ${\Delta}V$ as indicated corresponds to a 0 to 1 transition, to which we refer as $V_{\pi}$ when the chip length $L$ is given.}
    \label{fig:3}
\end{figure}

\subsection*{Geometry design}
Our design is based on the thin film Lithium Niobate TFLN platform, which is poised to become a material choice for applications relying on the electro-optic (EO) effect, offering significant potential in various applications~\cite{Krasnokutska_2019, Zhu:21} and enabling ultra-low loss waveguides~\cite{Krasnokutska:18}. However, this design concept can be applied to any thin film EO materials, including Barium titanate~\cite{abel2016hybrid,karvounis2020barium}.

Fig~\ref{fig:2} (c) shows the schematic of the proposed $\Delta{C}$-TDC based on Z-cut TFLN. Fig~\ref{fig:2} (d) and (e) present its cross-section of the coupled region and single waveguide, respectively. The TDC consists of two waveguides placed close to each other, allowing for optical power transfer via their evanescent field.

A $\Delta{C}$-TDC exhibits tunability due to the variation of the coupling coefficient in the coupled region (Fig~\ref{fig:3} (a) and (b)) while ensuring that the odd mode is perfectly antisymmetric and the even mode perfectly symmetric, thereby resulting in equal propagation constants for both waveguides~\cite{doi:https://doi.org/10.1002/0471213748.ch7}. To achieve this uniformity in propagation constants, it is imperative to maintain identical refractive index profiles for the coupled waveguides.

In order to utilize the strongest EO coefficient of LN while preserving uniformity in refractive index profiles, electrodes are placed on the bottom and top of the coupled region of the TDC. Both electrodes are much wider than the total width of the waveguides, generating a homogeneous electric field across two waveguides in the coupled region solely along the $z$ direction.

The relationship between the refractive index and the applied electric field in LN is expressed as:
\begin{align}
    \Delta(1/n^2)_{ij}={\sum_k}r_{ijk}E_k
    \label{eq:1}
\end{align}
where ${\Delta}(1/n^2)_{ij}$ is the change of the relative permittivity tensor, E$_k$ is the electric field vector, and $r_{ijk}$ is the EO tensor with $i, j, k$ corresponding to $x, y, z$ crystal coordinate. The tensor $r_{ijk}$ can further be expressed using two indices $r_{I,k}$ due to the symmetry property of LN as~\cite{Weis_1985,Zhu:21}:
\begin{align}
    r_{Ik}=
    \begin{pmatrix}
        0 & -r_{22} & r_{13}\\
        0 & r_{22} & r_{13}\\
        0 & 0 & r_{33}\\
        0 & r_{42} & 0\\
        r_{42} & 0 & 0\\
        -r_{22} & 0 & 0\\
    \end{pmatrix} \text{with } I=  \begin{cases}
      1, \quad ij=xx \\
      2, \quad ij=yy \\
      3, \quad ij=zz\\
      4, \quad ij=yz,zy\\
      5, \quad ij=xz,zx\\
      6, \quad ij=xy,yx
    \end{cases}.
    \label{eq:eo}
\end{align}
where $r_{13}\approx9.6$~pm/V, $r_{33}\approx30.9$~pm/V~\cite{electro-optic-coefficients}.

The change of the refractive index is given by the electro-optic effect (Eq~\ref{eq:1} and~\ref{eq:eo}), approximately ${\Delta}n_{x}{\approx}-1/2n_{o}^{3}r_{13}E_{z}$, ${\Delta}n_{y}{\approx}-1/2n_{o}^{3}r_{13}E_{z}$, ${\Delta}n_{z}{\approx}-1/2n_{e}^{3}r_{33}E_{z}$ based on first order Taylor expansion~\cite{Zhu:21}.
Subscripts \textit{o/e} stand for ordinary/extraordinary axis and the electric field across the two electrodes along $z$ direction is determined by COMSOL simulations.

\subsection*{Performance optimization}
In this section, we describe the details of optimizing the tunability performance of the proposed TDC design.

The TFLN refractive index is based on the Sellmeier equation (see supplementary material) provided by Nanoln, which is in line with our previously fabricated devices~\cite{Krasnokutska:18,Krasnokutska:19,Krasnokutska_2019,Krasnokutska:19_2}. We solve optical modes~\cite{AUTOKEY_modesolverpy} based on the electric-field-dependent (generated by voltage in a range $V\in\Delta{V}$) refractive index profile to calculate the effective refractive indices of the symmetric and the antisymmetric modes $N_s(V,\lambda)$ and $N_a(V,\lambda)$. The crossover length is calculated as
\begin{align}
    L_{c}(V,\lambda)=\frac{\lambda}{2(N_s(V,\lambda)-N_a(V,\lambda))}
    \label{eq:lc}
\end{align}
Where $\lambda$ is the wavelength of the operating light.
The coupling coefficient is given by
\begin{align}
    C(V,\lambda)=\frac{\pi}{2{\cdot}L_c(V,\lambda)}.
    \label{eq:c}
\end{align}
We construct a 2$\times$2 Hamiltonian $H(V,\lambda)$ for the TDC by applying continuous quantum walk model for DCs~\cite{PhysRevLett.102.253904, doi:10.1126/science.1193515}
\begin{align}
    H(V,\lambda) &= \begin{pmatrix} \beta_l(V,\lambda) & C(V,\lambda) \\ C(V,\lambda) & \beta_r(V,\lambda)  \end{pmatrix} 
\end{align}
where the waveguide propagation constants are assumed to be identical with their values derived from the supermodes $\beta_l(V,\lambda)$=$\beta_r(V,\lambda)$~\cite{Kapon:84}.

The unitary transformation implemented by the $\Delta{C}$-TDC is given by the time evolution of the Hamiltonian $H(V,\lambda)$ over the effective coupled length $L$
\begin{align}
    U(V,\lambda,L)=e^{-iH(V,\lambda)L}.
\end{align}
The unitary transformation implemented by a DC can be written in the form of~\cite{Matthews_2009}
\begin{align}
    U(V,\lambda,L) &= \begin{pmatrix} \sqrt{\eta(V,\lambda,L)} & i\sqrt{1-\eta(V,\lambda,L)} \\ i\sqrt{1-\eta(V,\lambda,L)} & \sqrt{\eta(V,\lambda,L)}  \end{pmatrix} 
\end{align}
where $\eta(V,\lambda,L)$ is the reflectivity. Accordingly, the reflectivity is given by
\begin{align}
    \eta(V,\lambda,L)=|U_{11}(V,\lambda,L)|^2.
\end{align}

We define $L_{\pi}$ as the minimum length that the TDC needs to implement a $0-1$ full reflectivity change within the range ${\Delta}V$ for a given wavelength $\lambda$. The requirement is then that 
${L_{\pi}C(V)}$ covers the full range between $k\pi$ and $\left(k+\frac{1}{2}\right)\pi$ or alternatively between $\left(k+\frac{1}{2}\right){\pi}$ and $(k+1)\pi$, for some $k{\in}\mathbb{Z}$. (Details can be found in the supplementary materials).

To determine the tunability performance of a $\Delta{C}$-TDC, we define as the figure of merit (FOM) for the tunability of a $\Delta{C}$-TDC:
\begin{align}
     \text{FOM}(\lambda)=\Delta{V}{\cdot}L_{\pi}
     \label{eq:measure}
 \end{align}
with a unit of ~V$\cdot$cm. The FOM indicates the resources of voltage range and coupling length required to achieve 0-to-1 reflectivity. Accordingly, it pinpoints an optimal combination in terms of the resources coupling length $L_{\pi}$ and voltage range $\Delta{V}$. This concept is similar to $V_{\pi}{\cdot}L$ in modulators, which indicates device size and modulation efficiency. We use $\Delta{V}{\cdot}L_{\pi}$ to characterize $\Delta{C}$-TDCs, as their tuning principle differs.

\begin{figure}
    \centering
    \includegraphics[width=0.5\columnwidth]{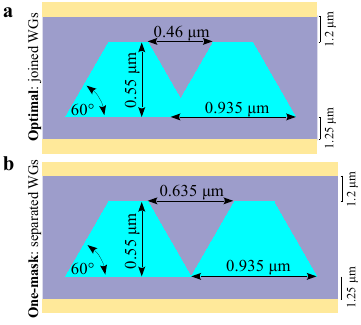}
    \caption{\textbf{Geometry parameters of proposed TFLN-based $\Delta{C}$-TDCs.} (\textbf{a}) The geometry with optimal FOM. (\textbf{b}) The optimal geometry allowing for one-mask fabrication.}
    \label{fig:4}
\end{figure}

\subsection*{Proposed devices}
In this section, we present simulation results for the proposed $\Delta{C}$-TDCs. The first coupler design corresponds to the fully optimized structure, which has a small waveguide gap of 0.46~$\mu$m (FOM=15.7~V$\cdot$cm at wavelength 1.55~$\mu$m) as shown in Fig~\ref{fig:4} (a) and requires two iterations of electron beam lithography (EBL) and etching. To simplify the fabrication process, we also provide the results for the second design, featuring a gap of 0.635~$\mu$m (FOM=19.5~V$\cdot$cm at wavelength 1.55~$\mu$m) as shown in Fig~\ref{fig:4} (b), which requires a single EBL and etching iteration. The modes and propagation simulations of the two coupler designs are reported in the supplementary material. 

\begin{figure}
    \centering
    \includegraphics[width=0.8\columnwidth]{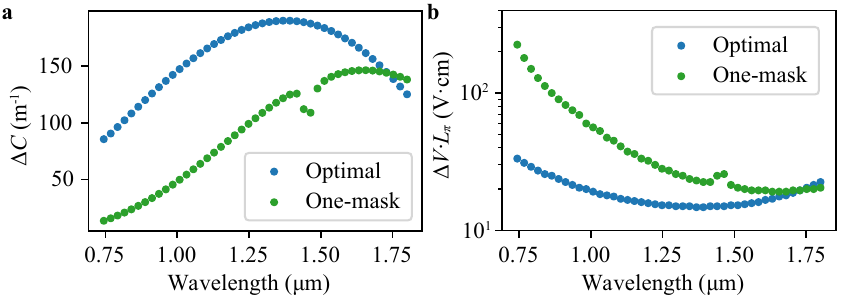}
    \caption{\textbf{Tunability of the two proposed $\Delta{C}$-TDCs.} (\textbf{a}) Coupling coefficient variation ${\Delta}C$ induced by $\Delta{V}=18$~V. (\textbf{b}) Tunability FOM at different wavelengths. At wavelengths of 1.4~$\mu$m and 1.6~$\mu$m, the optimal and one mask designs, respectively, require the least resources. (The out-of-order data points around 1.45~$\mu$m in the plots for one-mask design are the results of hybridization between the fundamental odd TE mode and fundamental even TM mode, which is expected to affect the coupling behavior at this wavelength.)}
    \label{fig:5}
\end{figure}

In the optimization process~\cite{kraft1988software,AUTOKEY_scipy_minimize}, we minimize the FOM of the $\Delta{C}$-TDC. We set the simulation in a voltage range $\Delta{V}=18$~V (a random but realistic value) at a wavelength of 1.55~$\mu$m and distance between the top and bottom thin film gold~\cite{Yakubovsky:17} electrodes 1.5~$\mu$m. The electrodes need to be placed with a minimum gap of 1~$\mu$m between waveguides to avoid optical losses due to absorption caused by electrodes. The parameters we optimized are thin film thickness, etching depth, waveguide top width, gap distance between the top of two waveguides, and sidewall angle in the range of $[0.3, 0.9]$~$\mu$m, $[0.1, 0.9]$~$\mu$m, $[0.1, 2.2]$~$\mu$m,$[0.1, 1.2]$~$\mu$m, and $[45^\circ,75^\circ]$ respectively.

The tunability depends on the relative magnitude of the static coupling coefficient and the variation of the coupling coefficient introduced within ${\Delta}V$. A greater static coupling coefficient requires a shorter coupled length to achieve the same tunability given by a fixed $\Delta{C}$. We report the variation of coupling coefficients of the two $\Delta{C}$-TDCs in the voltage range ${\Delta}V$ at different wavelengths in Fig~\ref{fig:5} (a). When $C\gg\Delta{C}$, the tunability can be considered as solely given by the variation in coupling coefficient within ${\Delta}V$.

Additionally, when the gap between two waveguides becomes very small, it is necessary to consider the effects of higher-order modes, including radiation modes, as they can introduce losses and limit the efficiency of power transfer. However, our waveguide structure is designed to support only the propagation of fundamental modes (single-mode waveguide), and the gap is not small enough for the two waveguides to behave as a single waveguide. Therefore, the impact of higher-order modes on the overall power transfer ratio is negligible for our designs, which is verified in 3D Finite-difference time-domain (FDTD) simulations, as shown in the supplementary materials.

We report the FOM of tunability of the two proposed $\Delta{C}$-TDCs in Fig~\ref{fig:5} (b). The FOM of the optimal design can achieve 15~V$\cdot$cm across wavelength from 1.25 to 1.58~$\mu$m and the FOM of the one-mask design is below 20.5~V$\cdot$cm across wavelength from 1.45 to 1.8~$\mu$m. The operation wavelength bandwidth is continuous and much broader than MZIs and $\Delta\beta$-TDCs. The optimal design achieves its best FOM at a wavelength of approximately 1.368~$\mu$m, with a 3-dB bandwidth (defined as the operational spectrum range achieving an FOM within less than a 50\% increase from its peak FOM at a given wavelength) ranging from 0.912~$\mu$m to 1.776~$\mu$m. The one-mask design achieves its best FOM at a wavelength of around 1.632~$\mu$m, with its 3-dB bandwidth starting at 1.248~$\mu$m.

For comparison, conventional $\Delta{\beta}$-TDCs, fabricated with titanium-diffused bulky LN technology, typically operate at switching voltages below 10~V but with a device length longer than 1~cm~\cite{doi:https://doi.org/10.1002/0471213748.ch18}. Furthermore, MZI modulators and $\Delta{\beta}$-TDC in which two waveguides experience electric fields in an opposite direction require less driving or electrode length to achieve $\pi$ phase change, particularly in x-cut LN~\cite{826874}. Specifically,  MZI modulators based on x-cut TFLN achieve $V_{\pi}{\cdot}L$ values of 1.8-3.1~V${\cdot}$cm in a monolithic manner~\cite{rao2016high,Wang_2018,wang2018nanophotonic}, and 2.1-3~V${\cdot}$cm in a hybrid manner~\cite{he2019high,7350128,ahmed2020subvolt} respectively. Additionally, $V_{\pi}{\cdot}L$ value of 7.4~V${\cdot}$cm has been demonstrated in z-cut TFLN with one arm modulation~\cite{hassanien2022compact}.

\begin{figure}
    \centering
    \includegraphics[width=0.8\columnwidth]{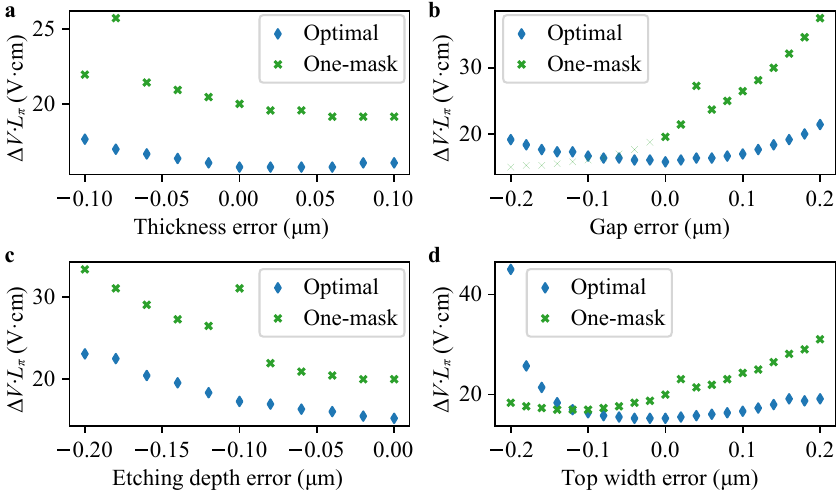}
    \caption{\textbf{Fabrication error analysis at 1.55~$\mu$m wavelength.} (\textbf{a}) Tunability FOM plot for thin film thickness error. (\textbf{b}) Tunability FOM plot for thin gap distance error. (\textbf{c}) Tunability FOM plot for etching depth error. (\textbf{d}) Tunability FOM plot for waveguide top width error.}
    \label{fig:6}
\end{figure}

\subsection*{Fabrication error analysis}
TFLN is a relatively new technology, and accordingly, the fabrication process is somewhat immature~\cite{Zhu:21}.
We analyze how fabrication errors affect the tunability of our couplers with respect to film thickness, gap distance, etching depth, and waveguide top width at a wavelength of 1.55~$\mu$m as shown in Fig~\ref{fig:6} (a), (b), (c) and (d) respectively, with a fixed 60$^\circ$ sidewall angle. The sidewall angle is given by the chemistry settings of the etching process and can be kept constant.
We postulate a uniform fabrication quality across both waveguides within the coupler structure while addressing individual error sources sequentially.
The simulation results indicate that a $\sim$100~nm fabrication error does not significantly affect the tunability and larger errors ($\sim$200~nm) can be compensated by increasing the coupler length or operating voltage range.

In addition to symmetric fabrication errors discussed above, asymmetric fabrication errors may also occur in practice. Such errors can result in dimensional differences between waveguides, preventing the device from achieving 0-1/1-0 reflectivity due to phase mismatches~\cite{doi:https://doi.org/10.1002/0471213748.ch7}. This challenge can be mitigated by improving fabrication quality or by introducing additional tuners to locally adjust the propagation constants of individual waveguides, similar to the dual-drive design~\cite{Perez-Lopez:19}. This approach can be feasible in electro-optic devices, as electro-optic effects offer precise control over local parameters~\cite{Yang_2024,yang2024programmable_single_qubit}.

\section*{Discussion}

In this paper, we proposed a design of a TDC with optimized geometries in the Z-cut TFLN platform. It allows arbitrary splitting ratio tuning while ensuring a common phase between waveguides across an extra-wide range of operating wavelengths. Furthermore, it exhibits intrinsic robustness to fabrication errors.

\begin{table}[h!]
\centering
\begin{tabular}{ |p{4cm}|p{1.22cm}|p{1.22cm}|p{1.22cm}|  }
 \hline
 Device & MZI & $\Delta\beta$-DC & $\Delta{C}$-DC\\
 \hline
 Fab-error robustness   & No    & No & Yes\\
  \hline
 Operation bandwidth & N.A.$^\text{a}$ & N.A.$^\text{a}$ & Wide\\
  \hline
  Phase conservation &   Yes  & No & Yes  \\
  \hline
   Modulation efficiency \newline @1.55~$\mu$m (V${\cdot}$cm) & \mline{1.8-3.1} & \mline{1.8-3.1}& \mline{15.7} \\
 \hline
   Number of bending section   & 4 &  2  & 2 \\
  \hline
\end{tabular}
\caption{Comparison of fundamental reflectivity tuning devices. ($^\text{a}$ MZI and $\Delta\beta$-DC can only operate at specific wavelengths in a periodic manner that allow them to achieve full 0-1 reflectivity, whereas $\Delta{C}$-DCs have continuous wide operational wavelength bandwidth.)}
\label{table:1}
\end{table}

Several crucial aspects need to be considered when comparing the design of the three fundamental reflectivity-tuning devices. A summarized comparison of fundamental tuning devices is presented in Table~\ref{table:1}. In terms of fabrication error, MZIs are well-known for being sensitive because of the requirement of perfect balanced directional couplers to achieve a full 0-1 reflectivity. Additionally, directional couplers are designed for specific operating wavelengths, rendering MZIs wavelength-dependent devices. On the other hand, $\Delta{\beta}$-TDCs have to be designed to present 0 reflectivity when static to ensure a complete 0-1/1-0 reflectivity, which makes such devices wavelength-dependent and less resilient to fabrication errors. On the contrary, $\Delta{C}$-TDCs can achieve a complete 0-1/1-0 reflectivity in a wide range of wavelengths and have shown robustness to fabrication error in simulations. Regarding the independent phase control required in applications such as quantum information processing, all three devices need an extra phase shifter. MZIs and $\Delta{C}$-TDCs can maintain a constant phase when working as reflectivity tuning devices, resulting in straightforward phase control. However, the phase of a $\Delta{\beta}$-TDC changes while its reflectivity is tuned and requires device-dependent characterization, which is challenging in practice. As for the driving voltage, $\Delta{C}$-TDCs require higher amplitudes compared to MZIs and $\Delta{\beta}$-TDCs when considering the same device size given the state-of-the-art technologies~\cite{doi:https://doi.org/10.1002/0471213748.ch18,Wang_2018}. This is because propagation constants tuning is more efficient for MZIs and $\Delta{\beta}$-TDCs. However, the driving voltage for $\Delta{C}$-TDCs can be reduced by increasing the length of the coupled region. Alternatively, the figure of merit (FOM) can be improved by decreasing the gap between electrodes or by utilizing materials with stronger electro-optic effects, potentially in a hybrid fashion~\cite{czornomaz2022bto}. Finally, in terms of the number of required bending sections, an MZI needs two more sections than both types of TDCs. This causes more losses, which becomes problematic when cascading multiple stages of MZIs~\cite{Miller_2015, Suzuki:15, Wang:20_3mzi}. Such cascading designs aim to overcome the challenges associated with the need for ideal balanced DCs in the MZIs. TDCs require fewer bendings, which makes them potentially experience less photon loss in large-scale circuits~\cite{Reck_1994, Clements_2016, Perez_2017}. However, they may need to be designed longer to compensate for fabrication errors and high driving voltage requirements.

In addition to Z-cut TFLN, our design can be implemented with X-cut TFLN, although more challenging. This is due to the requirements of the electrode positioning~\cite{Zhang:19} and the need for a small waveguide geometry that can only be realized with cutting-edge thin film technology. While nanofabrication technology in LNOI has facilitated the development of numerous wafer-scale devices, it may not be sufficient to meet the requirements for such devices. Moreover, based on state-of-the-art studies~\cite{doi:10.1126/science.aab3642, Miller_2015, Burgwal:17, Krasnokutska:18, Bao_2023}, it is reasonable to assume that each bending section contributes at least 0.1 dB loss. Additionally, TFLN waveguides with polished surface roughness have demonstrated propagation loss as low as 0.027~dB/cm~\cite{wu2018long}. Given that the $\Delta{C}$-TDC has one fewer bending section, a device no longer than approximately 3.7 cm with polished waveguides would have a loss advantage over MZIs that contain two bending sections. Enhancing sidewall roughness as well as designing efficient mode size converters are required, as these improvements will contribute to minimizing optical losses due to the small patterning features and ensuring the practical viability of the proposed device.

Our design concept is versatile and can be applied to any thin film EO materials, including Barium titanate~\cite{abel2016hybrid,karvounis2020barium}. As for the silicon platform, which is currently very popular in building PICs, our design can also be applied with thermal-optic devices. In contrast to electro-optic devices, thermo-optic devices do not offer the capability of cryogenic temperature operation~\cite{7463458} and high-speed reconfiguration~\cite{Wang_2018}. To avoid thermal crosstalk, waveguides on thermal-optic circuits need to be spaced at a distance even with more advanced fabrication technology~\cite {10011218,https://doi.org/10.1002/lpor.202000024}. Meanwhile, electro-optic devices need less space in the vertical direction relative to the light propagation with lower power consumption~\cite {Yang_2024}. Alternatively, the plasma dispersion effect is extensively employed for high-speed phase modulation~\cite{reed2010silicon}. However, devices utilizing this effect experience greater optical losses with increased tuning efficiency~\cite{1073206}. This is non-ideal for applications that are sensitive to optical losses, including building large-scale photonic circuits.

Lastly, the proposed $\Delta{C}$-TDC device is suitable for applications that are insensitive to high driving voltage but require maintaining a fixed relative phase. Moreover, such devices may be used as a tunable wavelength multiplexer–demultiplexer~\cite{1396022} or be cascaded to enhance filtering performance due to its wavelength-reflectivity-dependent operation. Additionally, $\Delta{C}$-TDCs could expand the functionalities of field-programmable photonic arrays and find applications in self-calibrating programmable PICs~\cite{Perez:16, Perez:18, Perez-Lopez:19, capmany2020programmable,Xu_2022}.

\section*{Data availability}
All data generated or analysed during this study are included in this published article and its supplementary information files.

\section*{Acknowledgements}
AP acknowledges an RMIT University Vice-Chancellor’s Senior Research Fellowship and a Google Faculty Research Award. This work was supported by the Australian Government through the Australian Research Council under the Centre of Excellence scheme (No: CE170100012). The authors thank Mr. Leon Wang for assistance in the schematic rendering.

\section*{Author contributions}
Y.Y. conceived and designed the device. Y.Y. and H.H. carried out the simulations. Y.Y., T.W., A.K., and A.P. prepared the manuscript. Y.Y., T.W., A.K., and A.P. analyzed the data and discussed the results. A.P. directed the project. 

\section*{competing interests}
The authors declare no competing interests.


\begin{thebibliography}{10}
\urlstyle{rm}
\expandafter\ifx\csname url\endcsname\relax
  \def\url#1{\texttt{#1}}\fi
\expandafter\ifx\csname urlprefix\endcsname\relax\def\urlprefix{URL }\fi
\expandafter\ifx\csname doiprefix\endcsname\relax\def\doiprefix{DOI: }\fi
\providecommand{\bibinfo}[2]{#2}
\providecommand{\eprint}[2][]{\url{#2}}

\bibitem{chrostowski_hochberg_2015}
\bibinfo{author}{Chrostowski, L.} \& \bibinfo{author}{Hochberg, M.}
\newblock \emph{\bibinfo{title}{Silicon Photonics Design: From Devices to Systems}} (\bibinfo{publisher}{Cambridge University Press}, \bibinfo{year}{2015}).

\bibitem{InP_based_photonic}
\bibinfo{author}{Smit, M.}, \bibinfo{author}{Williams, K.} \& \bibinfo{author}{van~der Tol, J.}
\newblock \bibinfo{journal}{\bibinfo{title}{{Past, present, and future of InP-based photonic integration}}}.
\newblock {\emph{\JournalTitle{APL Photonics}}} \textbf{\bibinfo{volume}{4}}, \bibinfo{pages}{050901}, \doiprefix\url{https://doi.org/10.1063/1.5087862} (\bibinfo{year}{2019}).

\bibitem{Zhu:21}
\bibinfo{author}{Zhu, D.} \emph{et~al.}
\newblock \bibinfo{journal}{\bibinfo{title}{Integrated photonics on thin-film lithium niobate}}.
\newblock {\emph{\JournalTitle{Adv. Opt. Photon.}}} \textbf{\bibinfo{volume}{13}}, \bibinfo{pages}{242--352}, \doiprefix\url{https://doi.org/10.1364/AOP.411024} (\bibinfo{year}{2021}).

\bibitem{Bogaerts_2020}
\bibinfo{author}{Bogaerts, W.} \emph{et~al.}
\newblock \bibinfo{journal}{\bibinfo{title}{Programmable photonic circuits}}.
\newblock {\emph{\JournalTitle{Nature}}} \textbf{\bibinfo{volume}{586}}, \bibinfo{pages}{207--216}, \doiprefix\url{https://doi.org/10.1038/s41586-020-2764-0} (\bibinfo{year}{2020}).

\bibitem{Marpaung_2019}
\bibinfo{author}{Marpaung, D.}, \bibinfo{author}{Yao, J.} \& \bibinfo{author}{Capmany, J.}
\newblock \bibinfo{journal}{\bibinfo{title}{Integrated microwave photonics}}.
\newblock {\emph{\JournalTitle{Nature Photonics}}} \textbf{\bibinfo{volume}{13}}, \bibinfo{pages}{80--90}, \doiprefix\url{https://doi.org/10.1038/s41566-018-0310-5} (\bibinfo{year}{2019}).

\bibitem{Shen_2017}
\bibinfo{author}{Shen, Y.} \emph{et~al.}
\newblock \bibinfo{journal}{\bibinfo{title}{Deep learning with coherent nanophotonic circuits}}.
\newblock {\emph{\JournalTitle{Nature Photonics}}} \textbf{\bibinfo{volume}{11}}, \bibinfo{pages}{441--446}, \doiprefix\url{https://doi.org/10.1038/nphoton.2017.93} (\bibinfo{year}{2017}).

\bibitem{Harris:18}
\bibinfo{author}{Harris, N.~C.} \emph{et~al.}
\newblock \bibinfo{journal}{\bibinfo{title}{Linear programmable nanophotonic processors}}.
\newblock {\emph{\JournalTitle{Optica}}} \textbf{\bibinfo{volume}{5}}, \bibinfo{pages}{1623--1631}, \doiprefix\url{https://doi.org/10.1364/OPTICA.5.001623} (\bibinfo{year}{2018}).

\bibitem{Yang_2024}
\bibinfo{author}{Yang, Y.} \emph{et~al.}
\newblock \bibinfo{journal}{\bibinfo{title}{Programmable high-dimensional hamiltonian in a photonic waveguide array}}.
\newblock {\emph{\JournalTitle{Nature Communications}}} \textbf{\bibinfo{volume}{15}}, \doiprefix\url{https://doi.org/10.1038/s41467-023-44185-z} (\bibinfo{year}{2024}).

\bibitem{shadbolt2012generating}
\bibinfo{author}{Shadbolt, P.~J.} \emph{et~al.}
\newblock \bibinfo{journal}{\bibinfo{title}{Generating, manipulating and measuring entanglement and mixture with a reconfigurable photonic circuit}}.
\newblock {\emph{\JournalTitle{Nature Photonics}}} \textbf{\bibinfo{volume}{6}}, \bibinfo{pages}{45--49}, \doiprefix\url{https://doi.org/10.1038/nphoton.2011.283} (\bibinfo{year}{2012}).

\bibitem{doi:10.1126/science.aab3642}
\bibinfo{author}{Carolan, J.} \emph{et~al.}
\newblock \bibinfo{journal}{\bibinfo{title}{Universal linear optics}}.
\newblock {\emph{\JournalTitle{Science}}} \textbf{\bibinfo{volume}{349}}, \bibinfo{pages}{711--716}, \doiprefix\url{https://doi.org/10.1126/science.aab3642} (\bibinfo{year}{2015}).

\bibitem{Reck_1994}
\bibinfo{author}{Reck, M.}, \bibinfo{author}{Zeilinger, A.}, \bibinfo{author}{Bernstein, H.~J.} \& \bibinfo{author}{Bertani, P.}
\newblock \bibinfo{journal}{\bibinfo{title}{Experimental realization of any discrete unitary operator}}.
\newblock {\emph{\JournalTitle{Physical Review Letters}}} \textbf{\bibinfo{volume}{73}}, \bibinfo{pages}{58--61}, \doiprefix\url{https://doi.org/10.1103/physrevlett.73.58} (\bibinfo{year}{1994}).

\bibitem{Clements_2016}
\bibinfo{author}{Clements, W.~R.}, \bibinfo{author}{Humphreys, P.~C.}, \bibinfo{author}{Metcalf, B.~J.}, \bibinfo{author}{Kolthammer, W.~S.} \& \bibinfo{author}{Walsmley, I.~A.}
\newblock \bibinfo{journal}{\bibinfo{title}{Optimal design for universal multiport interferometers}}.
\newblock {\emph{\JournalTitle{Optica}}} \textbf{\bibinfo{volume}{3}}, \bibinfo{pages}{1460}, \doiprefix\url{https://doi.org/10.1364/optica.3.001460} (\bibinfo{year}{2016}).

\bibitem{Perez_2017}
\bibinfo{author}{Perez, D.} \emph{et~al.}
\newblock \bibinfo{journal}{\bibinfo{title}{Silicon photonics rectangular universal interferometer}}.
\newblock {\emph{\JournalTitle{Laser \& Photonics Reviews}}} \textbf{\bibinfo{volume}{11}}, \bibinfo{pages}{1700219}, \doiprefix\url{https://doi.org/10.1002/lpor.201700219} (\bibinfo{year}{2017}).

\bibitem{Harris_2017}
\bibinfo{author}{Harris, N.~C.} \emph{et~al.}
\newblock \bibinfo{journal}{\bibinfo{title}{Quantum transport simulations in a programmable nanophotonic processor}}.
\newblock {\emph{\JournalTitle{Nature Photonics}}} \textbf{\bibinfo{volume}{11}}, \bibinfo{pages}{447--452}, \doiprefix\url{https://doi.org/10.1038/nphoton.2017.95} (\bibinfo{year}{2017}).

\bibitem{doi:10.1126/science.aar7053}
\bibinfo{author}{Wang, J.} \emph{et~al.}
\newblock \bibinfo{journal}{\bibinfo{title}{Multidimensional quantum entanglement with large-scale integrated optics}}.
\newblock {\emph{\JournalTitle{Science}}} \textbf{\bibinfo{volume}{360}}, \bibinfo{pages}{285--291}, \doiprefix\url{https://doi.org/https://doi.org/10.1126/science.aar7053} (\bibinfo{year}{2018}).

\bibitem{Qiang_2018}
\bibinfo{author}{Qiang, X.} \emph{et~al.}
\newblock \bibinfo{journal}{\bibinfo{title}{Large-scale silicon quantum photonics implementing arbitrary two-qubit processing}}.
\newblock {\emph{\JournalTitle{Nature Photonics}}} \textbf{\bibinfo{volume}{12}}, \bibinfo{pages}{534--539}, \doiprefix\url{https://doi.org/10.1038/s41566-018-0236-y} (\bibinfo{year}{2018}).

\bibitem{Bao_2023}
\bibinfo{author}{Bao, J.} \emph{et~al.}
\newblock \bibinfo{journal}{\bibinfo{title}{Very-large-scale integrated quantum graph photonics}}.
\newblock {\emph{\JournalTitle{Nature Photonics}}} \doiprefix\url{https://doi.org/10.1038/s41566-023-01187-z} (\bibinfo{year}{2023}).

\bibitem{doi:https://doi.org/10.1002/9783527695805.ch5}
\emph{\bibinfo{title}{Basic Optical Components}}, chap.~\bibinfo{chapter}{5}, \bibinfo{pages}{139--198} (\bibinfo{year}{2019}).
\newblock \eprint{https://onlinelibrary.wiley.com/doi/pdf/10.1002/9783527695805.ch5}.

\bibitem{kok2007linear}
\bibinfo{author}{Kok, P.} \emph{et~al.}
\newblock \bibinfo{journal}{\bibinfo{title}{Linear optical quantum computing with photonic qubits}}.
\newblock {\emph{\JournalTitle{Reviews of modern physics}}} \textbf{\bibinfo{volume}{79}}, \bibinfo{pages}{135--174}, \doiprefix\url{https://doi.org/10.1103/RevModPhys.79.135} (\bibinfo{year}{2007}).

\bibitem{doi:https://doi.org/10.1002/0471213748.ch7}
\bibinfo{author}{SALEH, B. E.~A.} \& \bibinfo{author}{TEICH, M.~C.}
\newblock \emph{\bibinfo{title}{Guided-Wave Optics}}, chap.~\bibinfo{chapter}{7}, \bibinfo{pages}{238--271} (\bibinfo{publisher}{John Wiley \& Sons, Ltd}, \bibinfo{year}{1991}).

\bibitem{doi:https://doi.org/10.1002/0471213748.ch18}
\bibinfo{author}{SALEH, B. E.~A.} \& \bibinfo{author}{TEICH, M.~C.}
\newblock \emph{\bibinfo{title}{Electro-Optics}}, chap.~\bibinfo{chapter}{18}, \bibinfo{pages}{696--736} (\bibinfo{publisher}{John Wiley \& Sons, Ltd}, \bibinfo{year}{1991}).

\bibitem{Miller_2015}
\bibinfo{author}{Miller, D. A.~B.}
\newblock \bibinfo{journal}{\bibinfo{title}{Perfect optics with imperfect components}}.
\newblock {\emph{\JournalTitle{Optica}}} \textbf{\bibinfo{volume}{2}}, \bibinfo{pages}{747}, \doiprefix\url{https://doi.org/10.1364/optica.2.000747} (\bibinfo{year}{2015}).

\bibitem{Burgwal:17}
\bibinfo{author}{Burgwal, R.} \emph{et~al.}
\newblock \bibinfo{journal}{\bibinfo{title}{Using an imperfect photonic network to implement random unitaries}}.
\newblock {\emph{\JournalTitle{Opt. Express}}} \textbf{\bibinfo{volume}{25}}, \bibinfo{pages}{28236--28245}, \doiprefix\url{https://doi.org/10.1364/OE.25.028236} (\bibinfo{year}{2017}).

\bibitem{826874}
\bibinfo{author}{Wooten, E.} \emph{et~al.}
\newblock \bibinfo{journal}{\bibinfo{title}{A review of lithium niobate modulators for fiber-optic communications systems}}.
\newblock {\emph{\JournalTitle{IEEE Journal of Selected Topics in Quantum Electronics}}} \textbf{\bibinfo{volume}{6}}, \bibinfo{pages}{69--82}, \doiprefix\url{https://doi.org/10.1109/2944.826874} (\bibinfo{year}{2000}).

\bibitem{Okayama}
\bibinfo{author}{Okayama, H.}
\newblock \bibinfo{title}{Lithium niobate electro-optic switching}.
\newblock In \emph{\bibinfo{booktitle}{Optical Switching}}, \bibinfo{pages}{39--81}, \doiprefix\url{https://doi.org/10.1007/0-387-29159-8_2} (\bibinfo{publisher}{Springer {US}}, \bibinfo{year}{2006}).

\bibitem{8351983}
\bibinfo{author}{Pérez-López, D.}, \bibinfo{author}{Sánchez, E.} \& \bibinfo{author}{Capmany, J.}
\newblock \bibinfo{journal}{\bibinfo{title}{Programmable true time delay lines using integrated waveguide meshes}}.
\newblock {\emph{\JournalTitle{Journal of Lightwave Technology}}} \textbf{\bibinfo{volume}{36}}, \bibinfo{pages}{4591--4601}, \doiprefix\url{https://doi.org/10.1109/JLT.2018.2831008} (\bibinfo{year}{2018}).

\bibitem{Perez-Lopez:19}
\bibinfo{author}{P\'{e}rez-L\'{o}pez, D.}, \bibinfo{author}{Gutierrez, A.~M.}, \bibinfo{author}{S\'{a}nchez, E.}, \bibinfo{author}{DasMahapatra, P.} \& \bibinfo{author}{Capmany, J.}
\newblock \bibinfo{journal}{\bibinfo{title}{Integrated photonic tunable basic units using dual-drive directional couplers}}.
\newblock {\emph{\JournalTitle{Opt. Express}}} \textbf{\bibinfo{volume}{27}}, \bibinfo{pages}{38071--38086}, \doiprefix\url{https://doi.org/10.1364/OE.27.038071} (\bibinfo{year}{2019}).

\bibitem{Krasnokutska_2019}
\bibinfo{author}{Krasnokutska, I.}, \bibinfo{author}{Tambasco, J.-L.~J.} \& \bibinfo{author}{Peruzzo, A.}
\newblock \bibinfo{journal}{\bibinfo{title}{Tunable large free spectral range microring resonators in lithium niobate on insulator}}.
\newblock {\emph{\JournalTitle{Scientific Reports}}} \textbf{\bibinfo{volume}{9}}, \doiprefix\url{https://doi.org/10.1038\%2Fs41598-019-47231-3} (\bibinfo{year}{2019}).

\bibitem{Krasnokutska:18}
\bibinfo{author}{Krasnokutska, I.}, \bibinfo{author}{Tambasco, J.-L.~J.}, \bibinfo{author}{Li, X.} \& \bibinfo{author}{Peruzzo, A.}
\newblock \bibinfo{journal}{\bibinfo{title}{Ultra-low loss photonic circuits in lithium niobate on insulator}}.
\newblock {\emph{\JournalTitle{Opt. Express}}} \textbf{\bibinfo{volume}{26}}, \bibinfo{pages}{897--904}, \doiprefix\url{https://doi.org/10.1364/OE.26.000897} (\bibinfo{year}{2018}).

\bibitem{abel2016hybrid}
\bibinfo{author}{Abel, S.} \emph{et~al.}
\newblock \bibinfo{journal}{\bibinfo{title}{A hybrid barium titanate--silicon photonics platform for ultraefficient electro-optic tuning}}.
\newblock {\emph{\JournalTitle{Journal of Lightwave Technology}}} \textbf{\bibinfo{volume}{34}}, \bibinfo{pages}{1688--1693}, \doiprefix\url{http://doi.org/10.1109/JLT.2015.2510282} (\bibinfo{year}{2016}).

\bibitem{karvounis2020barium}
\bibinfo{author}{Karvounis, A.}, \bibinfo{author}{Timpu, F.}, \bibinfo{author}{Vogler-Neuling, V.~V.}, \bibinfo{author}{Savo, R.} \& \bibinfo{author}{Grange, R.}
\newblock \bibinfo{journal}{\bibinfo{title}{Barium titanate nanostructures and thin films for photonics}}.
\newblock {\emph{\JournalTitle{Advanced Optical Materials}}} \textbf{\bibinfo{volume}{8}}, \bibinfo{pages}{2001249}, \doiprefix\url{https://doi.org/10.1002/adom.202001249} (\bibinfo{year}{2020}).

\bibitem{Weis_1985}
\bibinfo{author}{Weis, R.~S.} \& \bibinfo{author}{Gaylord, T.~K.}
\newblock \bibinfo{journal}{\bibinfo{title}{Lithium niobate: Summary of physical properties and crystal structure}}.
\newblock {\emph{\JournalTitle{Applied Physics A Solids and Surfaces}}} \textbf{\bibinfo{volume}{37}}, \bibinfo{pages}{191--203}, \doiprefix\url{https://doi.org/10.1007/bf00614817} (\bibinfo{year}{1985}).

\bibitem{electro-optic-coefficients}
\bibinfo{author}{Wong, K.}
\newblock \emph{\bibinfo{title}{8.4 Electro-Optic Coefficients of LiNbO3}}, \bibinfo{pages}{136} (\bibinfo{publisher}{Institution of Engineering and Technology (The IET)}, \bibinfo{year}{2002}).

\bibitem{Krasnokutska:19}
\bibinfo{author}{Krasnokutska, I.}, \bibinfo{author}{Tambasco, J.-L.~J.} \& \bibinfo{author}{Peruzzo, A.}
\newblock \bibinfo{journal}{\bibinfo{title}{Nanostructuring of lnoi for efficient edge coupling}}.
\newblock {\emph{\JournalTitle{Opt. Express}}} \textbf{\bibinfo{volume}{27}}, \bibinfo{pages}{16578--16585}, \doiprefix\url{https://doi.org/10.1364/OE.27.016578} (\bibinfo{year}{2019}).

\bibitem{Krasnokutska:19_2}
\bibinfo{author}{Krasnokutska, I.}, \bibinfo{author}{Chapman, R.~J.}, \bibinfo{author}{Tambasco, J.-L.~J.} \& \bibinfo{author}{Peruzzo, A.}
\newblock \bibinfo{journal}{\bibinfo{title}{High coupling efficiency grating couplers on lithium niobate on insulator}}.
\newblock {\emph{\JournalTitle{Opt. Express}}} \textbf{\bibinfo{volume}{27}}, \bibinfo{pages}{17681--17685}, \doiprefix\url{https://doi.org/10.1364/OE.27.017681} (\bibinfo{year}{2019}).

\bibitem{AUTOKEY_modesolverpy}
\bibinfo{title}{Github - jtambasco/modesolverpy: Photonic mode solver with a simple interface.}

\bibitem{PhysRevLett.102.253904}
\bibinfo{author}{Bromberg, Y.}, \bibinfo{author}{Lahini, Y.}, \bibinfo{author}{Morandotti, R.} \& \bibinfo{author}{Silberberg, Y.}
\newblock \bibinfo{journal}{\bibinfo{title}{Quantum and classical correlations in waveguide lattices}}.
\newblock {\emph{\JournalTitle{Phys. Rev. Lett.}}} \textbf{\bibinfo{volume}{102}}, \bibinfo{pages}{253904}, \doiprefix\url{https://doi.org/10.1103/PhysRevLett.102.253904} (\bibinfo{year}{2009}).

\bibitem{doi:10.1126/science.1193515}
\bibinfo{author}{Peruzzo, A.} \emph{et~al.}
\newblock \bibinfo{journal}{\bibinfo{title}{Quantum walks of correlated photons}}.
\newblock {\emph{\JournalTitle{Science}}} \textbf{\bibinfo{volume}{329}}, \bibinfo{pages}{1500--1503}, \doiprefix\url{https://doi.org/10.1126/science.1193515} (\bibinfo{year}{2010}).

\bibitem{Kapon:84}
\bibinfo{author}{Kapon, E.}, \bibinfo{author}{Katz, J.} \& \bibinfo{author}{Yariv, A.}
\newblock \bibinfo{journal}{\bibinfo{title}{Supermode analysis of phase-locked arrays of semiconductor lasers}}.
\newblock {\emph{\JournalTitle{Opt. Lett.}}} \textbf{\bibinfo{volume}{9}}, \bibinfo{pages}{125--127}, \doiprefix\url{https://doi.org/10.1364/OL.9.000125} (\bibinfo{year}{1984}).

\bibitem{Matthews_2009}
\bibinfo{author}{Matthews, J. C.~F.}, \bibinfo{author}{Politi, A.}, \bibinfo{author}{Stefanov, A.} \& \bibinfo{author}{O{\textquotesingle}Brien, J.~L.}
\newblock \bibinfo{journal}{\bibinfo{title}{Manipulation of multiphoton entanglement in waveguide quantum circuits}}.
\newblock {\emph{\JournalTitle{Nature Photonics}}} \textbf{\bibinfo{volume}{3}}, \bibinfo{pages}{346--350}, \doiprefix\url{https://doi.org/10.1038/nphoton.2009.93} (\bibinfo{year}{2009}).

\bibitem{kraft1988software}
\bibinfo{author}{Kraft, D.}
\newblock \bibinfo{journal}{\bibinfo{title}{A software package for sequential quadratic programming}}.
\newblock {\emph{\JournalTitle{Forschungsbericht- Deutsche Forschungs- und Versuchsanstalt fur Luft- und Raumfahrt}}}  (\bibinfo{year}{1988}).

\bibitem{AUTOKEY_scipy_minimize}
\bibinfo{title}{scipy.optimize.minimize — scipy v1.10.1 manual}.

\bibitem{Yakubovsky:17}
\bibinfo{author}{Yakubovsky, D.~I.}, \bibinfo{author}{Arsenin, A.~V.}, \bibinfo{author}{Stebunov, Y.~V.}, \bibinfo{author}{Fedyanin, D.~Y.} \& \bibinfo{author}{Volkov, V.~S.}
\newblock \bibinfo{journal}{\bibinfo{title}{Optical constants and structural properties of thin gold films}}.
\newblock {\emph{\JournalTitle{Opt. Express}}} \textbf{\bibinfo{volume}{25}}, \bibinfo{pages}{25574--25587}, \doiprefix\url{https://doi.org/10.1364/OE.25.025574} (\bibinfo{year}{2017}).

\bibitem{rao2016high}
\bibinfo{author}{Rao, A.} \emph{et~al.}
\newblock \bibinfo{journal}{\bibinfo{title}{High-performance and linear thin-film lithium niobate mach--zehnder modulators on silicon up to 50 ghz}}.
\newblock {\emph{\JournalTitle{Optics letters}}} \textbf{\bibinfo{volume}{41}}, \bibinfo{pages}{5700--5703}, \doiprefix\url{https://doi.org/10.1364/OL.41.005700} (\bibinfo{year}{2016}).

\bibitem{Wang_2018}
\bibinfo{author}{Wang, C.} \emph{et~al.}
\newblock \bibinfo{journal}{\bibinfo{title}{Integrated lithium niobate electro-optic modulators operating at {CMOS}-compatible voltages}}.
\newblock {\emph{\JournalTitle{Nature}}} \textbf{\bibinfo{volume}{562}}, \bibinfo{pages}{101--104}, \doiprefix\url{https://doi.org/10.1038/s41586-018-0551-y} (\bibinfo{year}{2018}).

\bibitem{wang2018nanophotonic}
\bibinfo{author}{Wang, C.}, \bibinfo{author}{Zhang, M.}, \bibinfo{author}{Stern, B.}, \bibinfo{author}{Lipson, M.} \& \bibinfo{author}{Lon{\v{c}}ar, M.}
\newblock \bibinfo{journal}{\bibinfo{title}{Nanophotonic lithium niobate electro-optic modulators}}.
\newblock {\emph{\JournalTitle{Optics express}}} \textbf{\bibinfo{volume}{26}}, \bibinfo{pages}{1547--1555}, \doiprefix\url{https://doi.org/10.1364/OE.26.001547} (\bibinfo{year}{2018}).

\bibitem{he2019high}
\bibinfo{author}{He, M.} \emph{et~al.}
\newblock \bibinfo{journal}{\bibinfo{title}{High-performance hybrid silicon and lithium niobate mach--zehnder modulators for 100 gbit s- 1 and beyond}}.
\newblock {\emph{\JournalTitle{Nature Photonics}}} \textbf{\bibinfo{volume}{13}}, \bibinfo{pages}{359--364}, \doiprefix\url{https://doi.org/10.1038/s41566-019-0378-6} (\bibinfo{year}{2019}).

\bibitem{7350128}
\bibinfo{author}{Jin, S.}, \bibinfo{author}{Xu, L.}, \bibinfo{author}{Zhang, H.} \& \bibinfo{author}{Li, Y.}
\newblock \bibinfo{journal}{\bibinfo{title}{Linbo3 thin-film modulators using silicon nitride surface ridge waveguides}}.
\newblock {\emph{\JournalTitle{IEEE Photonics Technology Letters}}} \textbf{\bibinfo{volume}{28}}, \bibinfo{pages}{736--739}, \doiprefix\url{https://doi.org/10.1109/LPT.2015.2507136} (\bibinfo{year}{2016}).

\bibitem{ahmed2020subvolt}
\bibinfo{author}{Ahmed, A. N.~R.} \emph{et~al.}
\newblock \bibinfo{journal}{\bibinfo{title}{Subvolt electro-optical modulator on thin-film lithium niobate and silicon nitride hybrid platform}}.
\newblock {\emph{\JournalTitle{Optics letters}}} \textbf{\bibinfo{volume}{45}}, \bibinfo{pages}{1112--1115}, \doiprefix\url{https://doi.org/10.1364/OL.381892} (\bibinfo{year}{2020}).

\bibitem{hassanien2022compact}
\bibinfo{author}{Hassanien, A.~E.}, \bibinfo{author}{Ghoname, A.~O.}, \bibinfo{author}{Chow, E.}, \bibinfo{author}{Goddard, L.~L.} \& \bibinfo{author}{Gong, S.}
\newblock \bibinfo{journal}{\bibinfo{title}{Compact mzi modulators on thin film z-cut lithium niobate}}.
\newblock {\emph{\JournalTitle{Optics Express}}} \textbf{\bibinfo{volume}{30}}, \bibinfo{pages}{4543--4552}, \doiprefix\url{https://doi.org/10.1364/OE.446665} (\bibinfo{year}{2022}).

\bibitem{yang2024programmable_single_qubit}
\bibinfo{author}{Yang, Y.} \emph{et~al.}
\newblock \bibinfo{journal}{\bibinfo{title}{Programmable quantum circuits in a large-scale photonic waveguide array}}.
\newblock {\emph{\JournalTitle{arXiv preprint arXiv:2405.13654}}} \doiprefix\url{https://doi.org/10.48550/arXiv.2405.13654} (\bibinfo{year}{2024}).

\bibitem{czornomaz2022bto}
\bibinfo{author}{Czornomaz, L.} \& \bibinfo{author}{Abel, S.}
\newblock \bibinfo{title}{Bto-enhanced silicon photonics--a scalable pic platform with ultra-efficient electro-optical modulation}.
\newblock In \emph{\bibinfo{booktitle}{Optical Fiber Communication Conference}}, \bibinfo{pages}{Th1J--1}, \doiprefix\url{https://doi.org/10.1364/OFC.2022.Th1J.1} (\bibinfo{year}{2022}).

\bibitem{Suzuki:15}
\bibinfo{author}{Suzuki, K.} \emph{et~al.}
\newblock \bibinfo{journal}{\bibinfo{title}{Ultra-high-extinction-ratio 2 {\texttimes} 2 silicon optical switch with variable splitter}}.
\newblock {\emph{\JournalTitle{Opt. Express}}} \textbf{\bibinfo{volume}{23}}, \bibinfo{pages}{9086--9092}, \doiprefix\url{https://doi.org/10.1364/OE.23.009086} (\bibinfo{year}{2015}).

\bibitem{Wang:20_3mzi}
\bibinfo{author}{Wang, M.}, \bibinfo{author}{Ribero, A.}, \bibinfo{author}{Xing, Y.} \& \bibinfo{author}{Bogaerts, W.}
\newblock \bibinfo{journal}{\bibinfo{title}{Tolerant, broadband tunable 2\text{$\times$}2 coupler circuit}}.
\newblock {\emph{\JournalTitle{Opt. Express}}} \textbf{\bibinfo{volume}{28}}, \bibinfo{pages}{5555--5566}, \doiprefix\url{https://doi.org/10.1364/OE.384018} (\bibinfo{year}{2020}).

\bibitem{Zhang:19}
\bibinfo{author}{Zhang, Q.} \emph{et~al.}
\newblock \bibinfo{journal}{\bibinfo{title}{Reconfigurable directional coupler in lithium niobate crystal fabricated by three-dimensional femtosecond laser focal field engineering}}.
\newblock {\emph{\JournalTitle{Photon. Res.}}} \textbf{\bibinfo{volume}{7}}, \bibinfo{pages}{503--507}, \doiprefix\url{https://doi.org/10.1364/PRJ.7.000503} (\bibinfo{year}{2019}).

\bibitem{wu2018long}
\bibinfo{author}{Wu, R.} \emph{et~al.}
\newblock \bibinfo{journal}{\bibinfo{title}{Long low-loss-litium niobate on insulator waveguides with sub-nanometer surface roughness}}.
\newblock {\emph{\JournalTitle{Nanomaterials}}} \textbf{\bibinfo{volume}{8}}, \bibinfo{pages}{910}, \doiprefix\url{https://doi.org/10.3390/nano8110910} (\bibinfo{year}{2018}).

\bibitem{7463458}
\bibinfo{author}{Elshaari, A.~W.}, \bibinfo{author}{Zadeh, I.~E.}, \bibinfo{author}{Jöns, K.~D.} \& \bibinfo{author}{Zwiller, V.}
\newblock \bibinfo{journal}{\bibinfo{title}{Thermo-optic characterization of silicon nitride resonators for cryogenic photonic circuits}}.
\newblock {\emph{\JournalTitle{IEEE Photonics Journal}}} \textbf{\bibinfo{volume}{8}}, \bibinfo{pages}{1--9}, \doiprefix\url{https://doi.org/10.1109/JPHOT.2016.2561622} (\bibinfo{year}{2016}).

\bibitem{10011218}
\bibinfo{author}{Prencipe, A.} \& \bibinfo{author}{Gallo, K.}
\newblock \bibinfo{journal}{\bibinfo{title}{Electro- and thermo-optics response of x-cut thin film linbo3 waveguides}}.
\newblock {\emph{\JournalTitle{IEEE Journal of Quantum Electronics}}} \textbf{\bibinfo{volume}{59}}, \bibinfo{pages}{1--8}, \doiprefix\url{https://doi.org/10.1109/JQE.2023.3234986} (\bibinfo{year}{2023}).

\bibitem{https://doi.org/10.1002/lpor.202000024}
\bibinfo{author}{Ceccarelli, F.} \emph{et~al.}
\newblock \bibinfo{journal}{\bibinfo{title}{Low power reconfigurability and reduced crosstalk in integrated photonic circuits fabricated by femtosecond laser micromachining}}.
\newblock {\emph{\JournalTitle{Laser \& Photonics Reviews}}} \textbf{\bibinfo{volume}{14}}, \bibinfo{pages}{2000024}, \doiprefix\url{https://doi.org/10.1002/lpor.202000024} (\bibinfo{year}{2020}).

\bibitem{reed2010silicon}
\bibinfo{author}{Reed, G.~T.}, \bibinfo{author}{Mashanovich, G.}, \bibinfo{author}{Gardes, F.~Y.} \& \bibinfo{author}{Thomson, D.~J.}
\newblock \bibinfo{journal}{\bibinfo{title}{Silicon optical modulators}}.
\newblock {\emph{\JournalTitle{Nature Photonics}}} \textbf{\bibinfo{volume}{4}}, \bibinfo{pages}{518--526}, \doiprefix\url{https://doi.org/10.1038/nphoton.2010.179} (\bibinfo{year}{2010}).

\bibitem{1073206}
\bibinfo{author}{Soref, R.} \& \bibinfo{author}{Bennett, B.}
\newblock \bibinfo{journal}{\bibinfo{title}{Electrooptical effects in silicon}}.
\newblock {\emph{\JournalTitle{IEEE Journal of Quantum Electronics}}} \textbf{\bibinfo{volume}{23}}, \bibinfo{pages}{123--129}, \doiprefix\url{https://doi.org/10.1109/JQE.1987.1073206} (\bibinfo{year}{1987}).

\bibitem{1396022}
\bibinfo{author}{Yamada, H.}, \bibinfo{author}{Chu, T.}, \bibinfo{author}{Ishida, S.} \& \bibinfo{author}{Arakawa, Y.}
\newblock \bibinfo{journal}{\bibinfo{title}{Optical directional coupler based on si-wire waveguides}}.
\newblock {\emph{\JournalTitle{IEEE Photonics Technology Letters}}} \textbf{\bibinfo{volume}{17}}, \bibinfo{pages}{585--587}, \doiprefix\url{https://doi.org/10.1109/LPT.2004.840926} (\bibinfo{year}{2005}).

\bibitem{Perez:16}
\bibinfo{author}{P\'{e}rez, D.}, \bibinfo{author}{Gasulla, I.}, \bibinfo{author}{Capmany, J.} \& \bibinfo{author}{Soref, R.~A.}
\newblock \bibinfo{journal}{\bibinfo{title}{Reconfigurable lattice mesh designs for programmable photonic processors}}.
\newblock {\emph{\JournalTitle{Opt. Express}}} \textbf{\bibinfo{volume}{24}}, \bibinfo{pages}{12093--12106}, \doiprefix\url{https://doi.org/10.1364/OE.24.012093} (\bibinfo{year}{2016}).

\bibitem{Perez:18}
\bibinfo{author}{P\'{e}rez, D.}, \bibinfo{author}{Gasulla, I.} \& \bibinfo{author}{Capmany, J.}
\newblock \bibinfo{journal}{\bibinfo{title}{Field-programmable photonic arrays}}.
\newblock {\emph{\JournalTitle{Opt. Express}}} \textbf{\bibinfo{volume}{26}}, \bibinfo{pages}{27265--27278}, \doiprefix\url{https://doi.org/10.1364/OE.26.027265} (\bibinfo{year}{2018}).

\bibitem{capmany2020programmable}
\bibinfo{author}{Capmany, J.} \& \bibinfo{author}{P{\'e}rez, D.}
\newblock \emph{\bibinfo{title}{Programmable integrated photonics}} (\bibinfo{publisher}{Oxford University Press}, \bibinfo{year}{2020}).

\bibitem{Xu_2022}
\bibinfo{author}{Xu, X.} \emph{et~al.}
\newblock \bibinfo{journal}{\bibinfo{title}{Self-calibrating programmable photonic integrated circuits}}.
\newblock {\emph{\JournalTitle{Nature Photonics}}} \textbf{\bibinfo{volume}{16}}, \bibinfo{pages}{595--602}, \doiprefix\url{https://doi.org/10.1038/s41566-022-01020-z} (\bibinfo{year}{2022}).

\end{thebibliography}
\end{document}


\title{Supplementary Information for \\A fixed phase tunable directional coupler based on coupling tuning}

\author{Yang Yang}
\address{Quantum Photonics Laboratory and Centre for Quantum Computation and Communication Technology, RMIT University, Melbourne, VIC 3000, Australia}

\author{Tim Weiss}
\address{Quantum Photonics Laboratory and Centre for Quantum Computation and Communication Technology, RMIT University, Melbourne, VIC 3000, Australia}

\author{Hamed Arianfard}
\address{Quantum Photonics Laboratory and Centre for Quantum Computation and Communication Technology, RMIT University, Melbourne, VIC 3000, Australia}

\author{Akram Youssry}
\address{Quantum Photonics Laboratory and Centre for Quantum Computation and Communication Technology, RMIT University, Melbourne, VIC 3000, Australia}

\author{Alberto Peruzzo}
\address{Quantum Photonics Laboratory and Centre for Quantum Computation and Communication Technology, RMIT University, Melbourne, VIC 3000, Australia}
\address{Qubit Pharmaceuticals, Advanced Research Department, Paris, France}

\maketitle
\subsection*{Phase comparison of fundamental tunable devices}
Mach-Zehnder interferometers (MZIs) require balanced (50:50) beam splitters, resulting in little fabrication error tolerance.
The conventional $\Delta\beta$-TDCs are wavelength and coupled length sensitive devices and require a perfect 0 reflectivity after fabrication with a zero relative phase between two waveguides (static devices).
We provide the numerical simulations via the time evolution method of an MZI and two types of TDC. The corresponding plots are reported in the main text.


An MZI is constructed by two balanced beam splitters and one phase shifter. Its unitary is given by

\begin{align}
    U_{\text{MZI}} &= U_{\text{50:50}}e^{i\theta\frac{\sigma_z}{2}}U_{\text{50:50}}\\
    &=\frac{1}{2}\begin{pmatrix} 1 & i \\ i & 1  \end{pmatrix}\begin{pmatrix} e^{i\frac{\theta}{2}} & 0 \\ 0 & e^{-i\frac{\theta}{2}}  \end{pmatrix}\begin{pmatrix} 1 & i \\ i & 1  \end{pmatrix}\\
    &=\begin{pmatrix} i\sin{\frac{\theta}{2}} & i\cos{\frac{\theta}{2}} \\ i\cos{\frac{\theta}{2}} & -i\sin{\frac{\theta}{2}}  \end{pmatrix},
\end{align}

where the relative phase difference is consistently equal to 0, except when $\theta=0$, the MZI implements a Pauli \textit{X} gate with an undefined relative phase. 



The Hamiltonian of a $\Delta\beta$-TDC (as shown in Fig~1, without considering the change of coupling coefficient) can be represented as
\begin{align}
    H_{\Delta\beta\text{-TDC}} &= \begin{pmatrix} \beta+\Delta\beta(V) & C \\
    C & \beta-\Delta\beta(V)  \end{pmatrix}, 
\end{align}
where $\Delta\beta(V)$ is the voltage-dependent propagation constant change. The unitary of such a $\Delta\beta$-TDC is given by the time evolution of its Hamiltonian $H$ over the effective coupled length $L$,
\begin{align}
    U_{\Delta\beta\text{-TDC}}=e^{-iH_{\Delta\beta\text{-TDC}}L}.
\end{align}

It can be shown that the angles of $U_{11}$ and $U_{12}$ are given by
\begin{align}
    \angle{U_{11}}&=\tan^{-1}\left(\frac{-\Delta\beta\tan{(\sqrt{C^2+\Delta\beta^2}L)}}{\sqrt{C^2+\Delta\beta^2}}\right)\\
    \angle{U_{12}}&=-\frac{\pi}{2}\text{sgn}\left(\frac{C\sin(\sqrt{C^2+\Delta\beta^2}L)}{\sqrt{C^2+\Delta\beta^2}} \right)
\end{align}

The relative phase is given by
\begin{align}
    \Delta\phi=\angle{U_{11}}-\angle{U_{12}}.
\end{align}

Therefore, $\angle{U_{11}}$ and $\Delta\phi$ changes with $\Delta\beta$, results in the absence of independent controllable phase.


The Hamiltonian of a $\Delta{C}$-TDC is given by
\begin{align}
    H_{\Delta{C}\text{-TDC}} &= \begin{pmatrix} \beta & C(V) \\
    C(V) & \beta  \end{pmatrix}, 
\end{align}
where $C(V)$ is the voltage-dependent coupling coefficient. In this case, we have $\Delta\beta=0$, which implies that

\begin{align}
    \angle{U_{11}}&=0,
\end{align}
and the relative phase given by a ${\Delta}C$-TDC can be represented as
\begin{align}
     \Delta\phi&=\begin{cases}
        -\frac{\pi}{2}, k\pi < CL { \le } (k+\frac{1}{2})\pi, k{\in}\mathbb{Z} \\
        \frac{\pi}{2}, (k+\frac{1}{2})\pi < {CL} { \le } (k+1)\pi, k{\in}\mathbb{Z}.
    \end{cases}
\end{align}


If we tune the coupling such that $\frac{\pi k}{L} < C  \le  \frac{(k + 1/2)\pi}{L}, k{\in}\mathbb{Z}$ or $\frac{(k + 1/2)\pi}{L} < C  \le  \frac{(k + 1)\pi}{L}, k{\in}\mathbb{Z}$, then the relative phase between the two arms is fixed.



\subsection{Optical modes and 3D FDTD simulations}
\begin{figure}
    \centering
    \includegraphics[width=1\columnwidth]{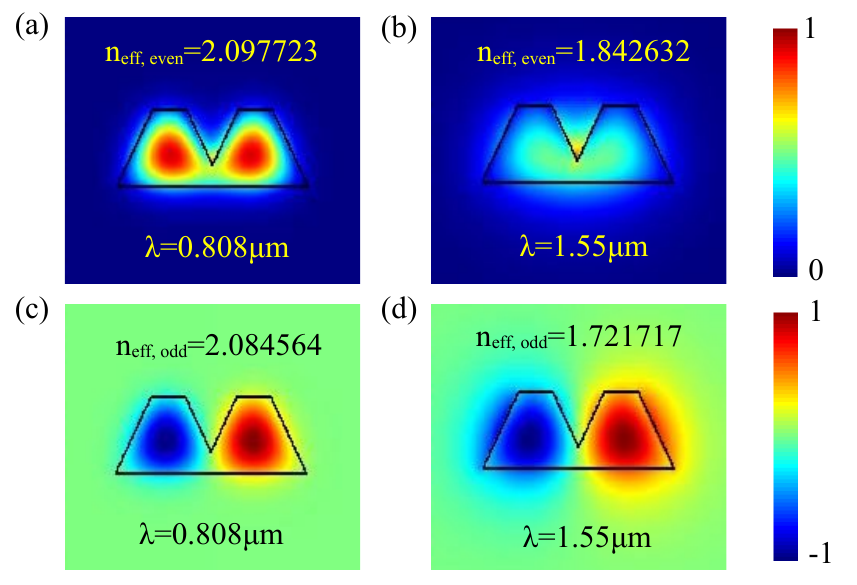}
    \caption{\textbf{Solved modes of device the optimal design.} The coupler geometry is represented by the black profile, while the electric field strength is indicated by the colorbar. (a) Even mode of at wavelength 0.808~$\mu$m. (b) Even mode of at wavelength 1.55~$\mu$m. (c) Odd mode of at wavelength 0.808~$\mu$m. (d) Odd mode of at wavelength 1.55~$\mu$m.}
    \label{fig:modes_geo1}
\end{figure}

\begin{figure}
    \centering
    \includegraphics[width=1\columnwidth]{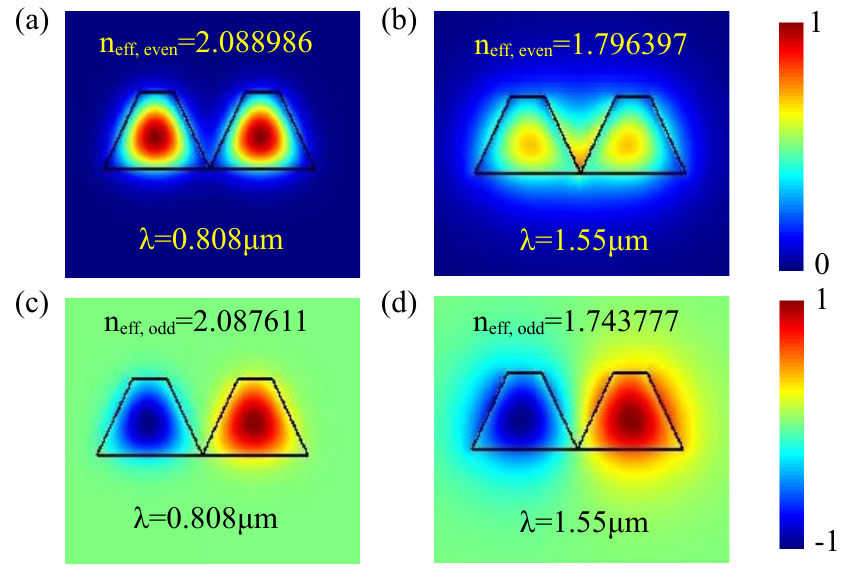}
    \caption{\textbf{Solved modes of the one-mask design.} The coupler geometry is represented by the black profile, while the electric field strength is indicated by the colorbar. (a) Even mode of at wavelength 0.808~$\mu$m. (b) Even mode of at wavelength 1.55~$\mu$m. (c) Odd mode of at wavelength 0.808~$\mu$m. (d) Odd mode of at wavelength 1.55~$\mu$m.}
    \label{fig:modes_geo2}
\end{figure}

\begin{figure}
    \centering
    \includegraphics[width=1\columnwidth]{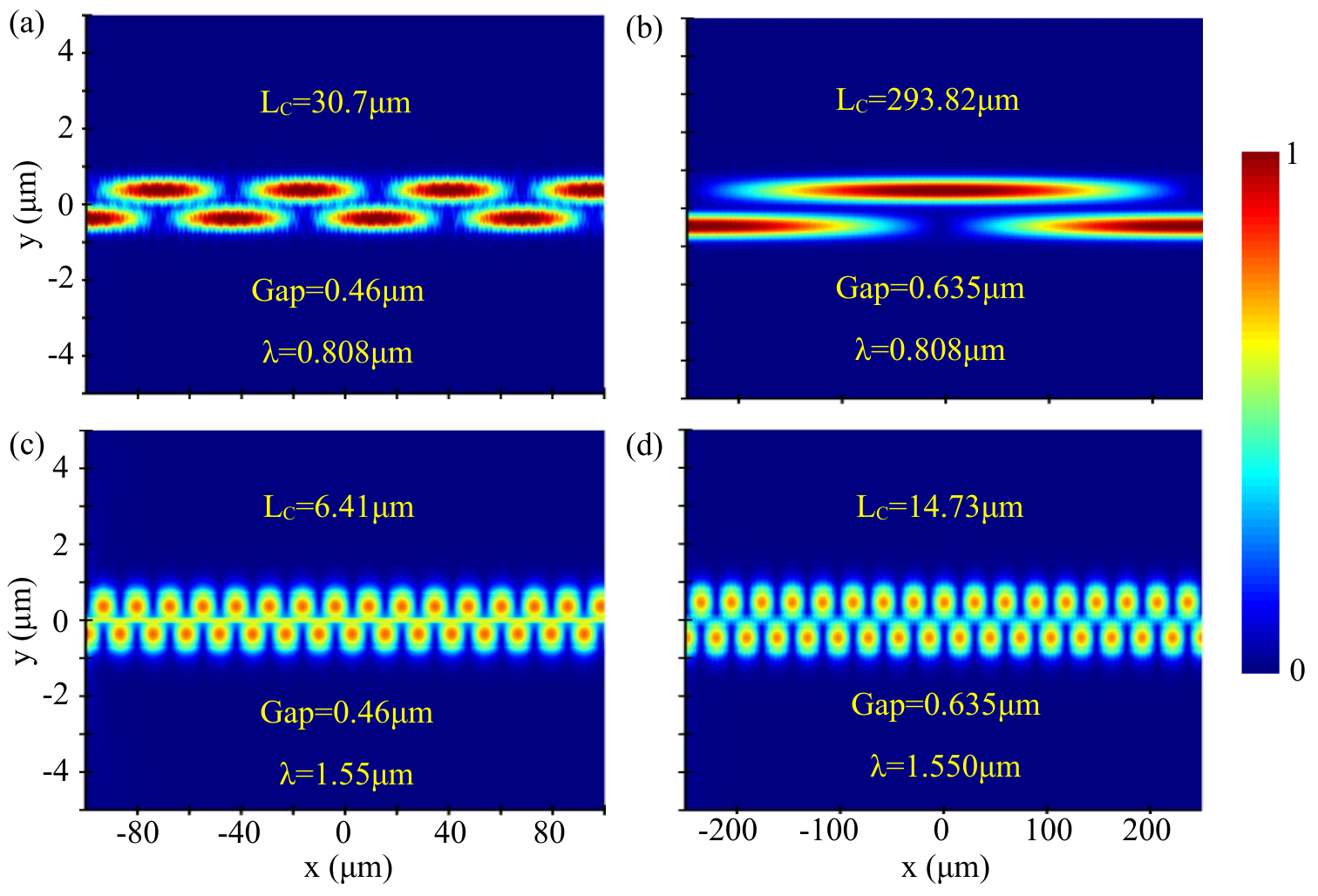}
    \caption{\textbf{3D FDTD propagation simulation of the coupled region.} (a) Simulation of the optimal design at wavelength 0.808~$\mu$m. (b) Simulation of one-mask design at wavelength 0.808~$\mu$m. (c) Simulation of the optimal design at wavelength 1.55~$\mu$m. (d) Simulation of one-mask design at wavelength 1.55~$\mu$m}
    \label{fig:propagation}
\end{figure}

In Fig~\ref{fig:modes_geo1} and \ref{fig:modes_geo2}, we report the solved TE modes as well as their effective refractive indices of the optimal and one-mask designs at wavelengths 0.808~$\mu$m and 1.55~$\mu$m when no voltage is applied. TM modes have less tunability than TE modes due to the crystal and electric field directions design of the device. 

In Fig~\ref{fig:propagation}, three-dimensional (3D) finite-difference time-domain (FDTD) simulations of the internal propagation of two couplers without tuning at wavelengths 0.808~$\mu$m and 1.55~$\mu$m are reported to verify the crossover/beat length $L_c$ (Eq~3) based on the solved modes.

\subsection{Sellmeier equation}
The Sellmeier equation used to calculate the refractive index of TFLN provided by Nanoln is given by
\begin{align}
    n_o(\lambda)^2=1+ \frac{2.7049\lambda^2}{\lambda^2-0.00868}+ \frac{1.1947\lambda^2}{\lambda^2-0.07231}+ \frac{12.614\lambda^2}{\lambda^2-474.6}
    \\
    n_e(\lambda)^2=1+ \frac{2.7939\lambda^2}{\lambda^2-0.01}+ \frac{0.8148\lambda^2}{\lambda^2-0.07075}+ \frac{8.9543\lambda^2}{\lambda^2-416.08}
 \end{align}
where subscript \textit{o/e} are ordinary/extraordinary axis and $\lambda$ is the wavelength.
\vspace{5cm}